# DETECTION OF DANGEROUS MAGNETIC FIELD RANGES FROM TABLETS BY CLUSTERING ANALYSIS

**Darko Brodić**
*Technical Faculty in Bor*
*University of Belgrade, Serbia*
dbrodic@tfbor.bg.ac.rs

**Alessia Amelio**
*DIMES*
*University of Calabria, Italy*
aamelio@dimes.unical.it

**Abstract**

*The paper considers the problem of the extremely low frequency magnetic field radiation generated by the tablet computers. Accordingly, the measurement of the magnetic field radiation from a set of tablets is carried out. Furthermore, the measurement results are analyzed and clustered according to the K-Medians algorithm to obtain different magnetic field ranges. The obtained cluster ranges are evaluated according to the reference level proposed by the TCO standard in order to define dangerous areas in the neighborhood of tablet, which are established during the typical work with tablet computers. Analysis shows that dangerous areas correspond to specific inner components of tablet, and gives suggestions to users for a safe usage of tablet and to companies producing tablet components for limiting the risk of magnetic field exposure.*

## INTRODUCTION

The humans are constantly exposed to various kinds of magnetic field radiation. Essentially, the source of the magnetic field dictates the characteristics of the magnetic field. The magnetic fields' sources can be natural or artificial. A natural sourced magnetic field is created by the Earth. It has constant amplitude and frequency. On the contrary, the artificial magnetic field varies all the time. Artificial magnetic field is a consequence of technology development. In its lower frequency part, it spreads over the Extremely Low Frequency (ELF), i.e. between 30 and 300 Hz.

Tablet is a highly portable mobile computer. It includes a touchscreen, battery, and motherboard with the included circuitry. Also, it is equipped with additional components like: microphone, loudspeaker, cameras, and a variety of sensors. Its screen is at least 7" wide. If it has no keyboard supplied with, then it is typically called booklet. Otherwise, it is called convertible tablet. It radiates a magnetic field like any other computer.

Many standards have proposed to use electronic devices in the safe way. One of the most spread is the TCO standard. It prescribed the measurement geometry and test procedure. It proposed that the safe limit of the ELF magnetic field radiation is bounded to 200 nT. The measurement geometry includes the measurement points at 0.30 m in front of and around the tablet computer [1]. Many researchers in their studies have pointed out the dangers of the emitted magnetic field to the human's health [2], [3]. Accordingly, they have proposed different safe limits of the ELF magnetic field radiation as 0.2 µT [3], 0.4 µT [2], and 1 µT [4]. In this paper, we use the reference proposed by TCO standard equal to 0.2 µT above which there is a danger to the human's health.

We explore the elements of the ELF magnetic field radiation initiated by tablet devices. Although some researchers explored the dangerous effect of laptop on human's health [5], [6], [7], to the best of our knowledge, nobody has explored the problem linked with the tablet computers yet. They come into the focus because of their common using characteristics:

(1) Wide-spreading,
(2) High portability, and
(3) Tendency to be used in close contact with the users' body.

Taking into account all aforementioned, the researching in this direction is of great importance to the human's health preservation.

In this paper, we take into account 6 different tablets in order to measure their ELF magnetic field radiation. Then, we cluster the magnetic field measuring values by K-Medians algorithm to designate magnetic field ranges associated to the dangerous areas of the tablet.

The paper is organized as follows. Section 2 explains the measurement procedure. Section 3 presents the result of the measurement and clustering. Section 4 discusses the obtained results. Section 5 draws conclusions.

**MEASUREMENT PROCEDURE**

**Magnetic field**. The magnetic field is established around its emitter. If it is uniform [5], then it is calculated as:

$$\mathbf{B(r)} = B_x \cdot \mathbf{x} + B_y \cdot \mathbf{y} + B_z \cdot \mathbf{z}, \quad (1)$$

where $\mathbf{x}$, $\mathbf{y}$ and $\mathbf{z}$ are the positional vectors, which are orthogonal to each other and $B_x$, $B_y$ and $B_z$ represent the magnitudes of the magnetic flux density in the direction of these vectors, respectively. The measuring devices measure the above scalar components of the magnetic flux density $B$. Hence, the root mean square (RMS) of the magnetic flux density $B$ can be calculated as:

$$B = \sqrt{(B_x^2 + B_y^2 + B_z^2)} \quad (2)$$

The experiment is carried out by measuring the ELF magnetic field on each tablet, i.e. at its top and bottom area in 9 different points. These points are referred as $tmp_i$ at the top part, and $bmp_i$ at the bottom part of tablet with $i=1...9$. Fig. 1 illustrates the tablet's top and bottom measuring points.

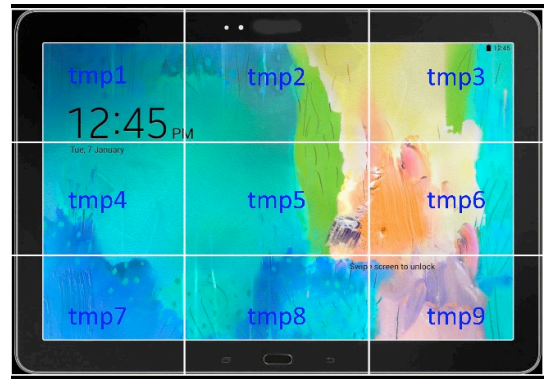

(a)

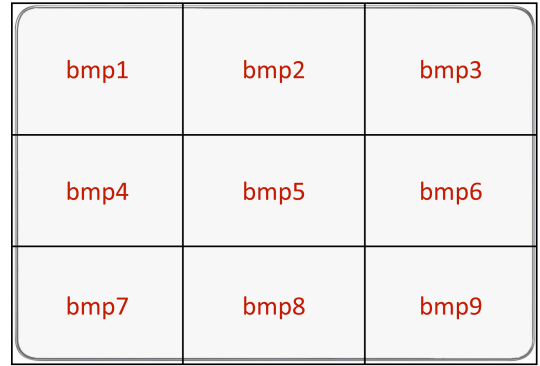

(b)

*Fig. 1. The tablet measuring points: (a) at the top part, (b) at the bottom part.*

**Measuring devices**. The ELF magnetic field is measured by measuring device Lutron EMF-828. Analysis is performed on 6 tablets made by different manufacturer, named as T1,...,T6. The first one, T1, has motherboard inside its body and it is equipped with external keyboard. The other ones, T2,...,T6, have also motherboard inside their body, but they don't have external keyboard. Keyboard of T1 is not considered, because it emits a negligible magnetic field radiation.

**Magnetic field radiation clustering**. Measuring points are subjected to clustering for automatic detection of the areas at high magnetic field emission on tablets. It is obtained by considering magnetic field radiation features for each point. Then, clustering is performed on points based on their feature representation. It realizes magnetic field ranges from which the high-risk areas can be effectively detected.

**Feature detection**. For each of the 6 tablets, each measuring point at top part, $tmp_i$ ($i=1...9$), and at bottom part, $bmp_i$ ($i=1...9$), is associated to a single feature. It represents the RMS of the measured magnetic flux density at that point, i.e. *B*. Consequently, each tablet determines 9 features at top part and 9 features at bottom part. Hence, two separate datasets are created from measuring points of tablets. The first one collects 54 features from top positions and is named as *top*. The second one contains 54 features from bottom positions and is named as *bottom*.

**Clustering algorithm.** We employ K-Medians algorithm for clustering of features from top and bottom parts of tablets [8]. It is a well-known representative of the center-based clustering methods, adopted in some contexts for clustering of one-dimensional data, like measurements [6], [9], [10]. The number of clusters K is fixed a priori and it is an input parameter of the method. Center-based characteristic derives from the concept of centroid, which is the representative of the cluster in the algorithm. For each cluster, its centroid is computed as the median value of the measuring point features belonging to that cluster. Its advantage relies on robustness to outliers and ability to generate more compact clusters than K-Means algorithm [11].

K-Medians runs in three main phases:
(1) The initial centroid selection,
(2) The measuring points' assignment, and
(3) The centroid re-computation.

In the first phase, the algorithm selects K initial centroids. It is a critical aspect, because this choice can influence the final clustering result. In the second phase, each measuring point is assigned to the centroid which is the nearest in terms of L1 distance between the corresponding features. In the last phase, the K centroids are re-computed from the newly obtained clusters. Phases (2) and (3) are iterated multiple times, until centroids don't modify anymore their position. Final centroids characterize the clustering solution of measuring points. Because the features are employed for clustering, final solution determines also ranges of magnetic field radiation, associated to tablet points.

**RESULTS AND DISCUSSION**

K-Medians is applied separately on top and bottom datasets. For each dataset, the K input parameter is fixed to 5, which demonstrated to be suitable for laptop analysis [6], [12] in order to obtain clusters of points associated to 5 magnetic field ranges. Table 1 shows the obtained ranges, $R_1,...,R_5$, corresponding to clusters of top (on the left) and bottom (on the right) datasets.

*Table. 1. Magnetic field ranges obtained for top and bottom datasets.*

| Top range | min-max ($\mu T$) | Bottom range | min-max ($\mu T$) |
|---|---|---|---|
| $R_1$ | 0.1965 - 0.8629 | $R_1$ | 0.8592 - 0.9571 |
| $R_2$ | 0.1204 - 0.1879 | $R_2$ | 0.2867 - 0.5101 |
| $R_3$ | 0.0768 - 0.0990 | $R_3$ | 0.1841 - 0.2627 |
| $R_4$ | 0.0412 - 0.0735 | $R_4$ | 0.0927 - 0.1530 |
| $R_5$ | 0.0100 - 0.0224 | $R_5$ | 0.0141 - 0.0849 |

Each range is defined in terms of minimum and maximum values of the measuring points associated to that range. Also, ranges are sorted in decreasing order (from highest to lowest) based on minimum and maximum values. It is worth to note that points emitting values above 0.2 µT are well-separated from points emitting values below the limit. It is fully observable for top part with $R_1$, strongly delimiting points above the limit. It is also approximately observable for bottom part, where points emit above the limit in $R_1$, $R_2$ and $R_3$. Another important observation is that bottom points of tablets determine the highest ranges of magnetic field radiation. In fact, the highest peak of 0.96 µT is reached in $R_1$ for bottom points. Also, the highest peak for top points is 0.86 µT, which is almost equal to the minimum of $R_1$ for bottom points. Furthermore, $R_3$, $R_4$ and $R_5$ from top points are merged into $R_5$, which is the safest range for bottom points. Again, three ranges at bottom part, $R_1$, $R_2$ and $R_3$, exhibit values above the reference limit of 0.2 µT. On the contrary, only $R_1$ for top points corresponds to values above 0.2 µT.

Ranges detected from top and bottom datasets are compared, extended to have the minimum of each range just above the maximum of the previous range, and aligned to create a unified classification of tablet points. The 7 classes are reported in Fig. 2.

**Fig. 2.** *The 7 classes obtained from top ranges and bottom ranges.*

| | | | TOP | BOTTOM |
|---|---|---|---|---|
| >= 0,2 uT | 1 | Highly dangerous | > 0,19 | > 0,85 |
| | 2 | Middle dangerous | - | 0,29 - 0,85 |
| | 3 | Low dangerous | - | 0,18 - 0,28 |
| < 0,2 uT | 4 | Low safe | 0,12 - 0,19 | 0,09 - 0,17 |
| | 5 | Low medium safe | 0,08 - 0,11 | - |
| | 6 | Medium safe | 0,04 - 0,07 | - |
| | 7 | Highly safe | 0,01 - 0,03 | 0,01 - 0,08 |

They are:
(1) Highly dangerous,
(2) Middle dangerous,
(3) Low dangerous,
(4) Low safe,
(5) Low medium safe,
(6) Medium safe, and
(7) Highly safe.

Classification is based on the top and bottom ranges approximately above and below the reference limit of 0.2 µT. In the case of top, we have that $R_2$, $R_3$, $R_4$ and $R_5$ are below 0.2 µT. For this reason, they are marked respectively as low safe, low medium safe, medium safe and highly safe. $R_1$, enveloping all the values above the reference limit, is marked as highly dangerous. In the case of bottom, only $R_4$ and $R_5$ are totally below 0.2 µT. Consequently, $R_5$ is marked as highly safe, while $R_4$, mostly overlapping with top $R_2$, is marked as low safe. Because $R_1$, $R_2$ and $R_3$ are above the reference limit, they are marked respectively as highly dangerous, middle dangerous and low dangerous. In particular, because $R_1$ exhibits the highest values for bottom, it is marked as highly dangerous.

Obtained classes are differently colored and numbered. Then, maps of dangerousness are built on the 6 tablet representations involved in clustering and reported in Fig. 3.

For each class, it shows the different points at top and bottom parts of tablets emitting values in that class (at top or at bottom). Also, the corresponding feature values are reported inside the top and bottom parts of tablet, differentiated according to the reference limit of 0.2 µT (in dark pink are values >=0.2 µT).

Looking at the features above the reference limit, we observe their association with the different tablet points and their distribution inside the different dangerousness classes. From the map, we observe that the highest peaks of highly dangerous level are located at points corresponding to CPU. It is mainly visible at top and bottom part of T1, with values of 0.52 µT and 0.93 µT, at top and bottom part of T5, with values of 0.86 µT and 0.96 µT, and at bottom part of T3, with a value of 0.86 µT. The points emitting at high level correspond also to the battery area. It is observable at bottom part of T2, emitting at low dangerous magnetic field level, with a peak in 0.20 µT, and at top parts of T3 and T6, emitting at highly dangerous level, with the highest peaks of 0.43 µT and 0.45 µT. Again, T4 emits at middle and low dangerous levels at bottom part in the area of CPU, with the highest peak of 0.51 µT.

K-Medians algorithm has been implemented in Matlab R2015a. Experiment has been performed on a notebook with CPU quad-core 2.2 GHz, 16 GB RAM and UNIX operating system. To overcome the initialization problem, K-Medians has been run 50 times with a new set of initial centroids, on each dataset. The result best minimizing the K-Medians function has been selected as the final clustering solution.

From this analysis, it is worth to note that tablets are characterized by magnetic field emission which can be classified based on different risk levels. From the obtained results, it is clear that CPU and battery areas emit at very high level of magnetic field radiation. Also, we found that the bottom part of tablet emits stronger magnetic field level than the top part. Consequently, users are strongly encouraged to not put tablet bottom on its knees, avoiding the influence of magnetic field radiation on skin and bones. Furthermore, it could be useful to employ a stylus pen for touchscreen, avoiding the direct contact of the fingers with tablet screen. On the other hand, companies producing tablets and inner components should seriously consider the risk, which is connected to magnetic field exposure and to make effort for designing low emission components.

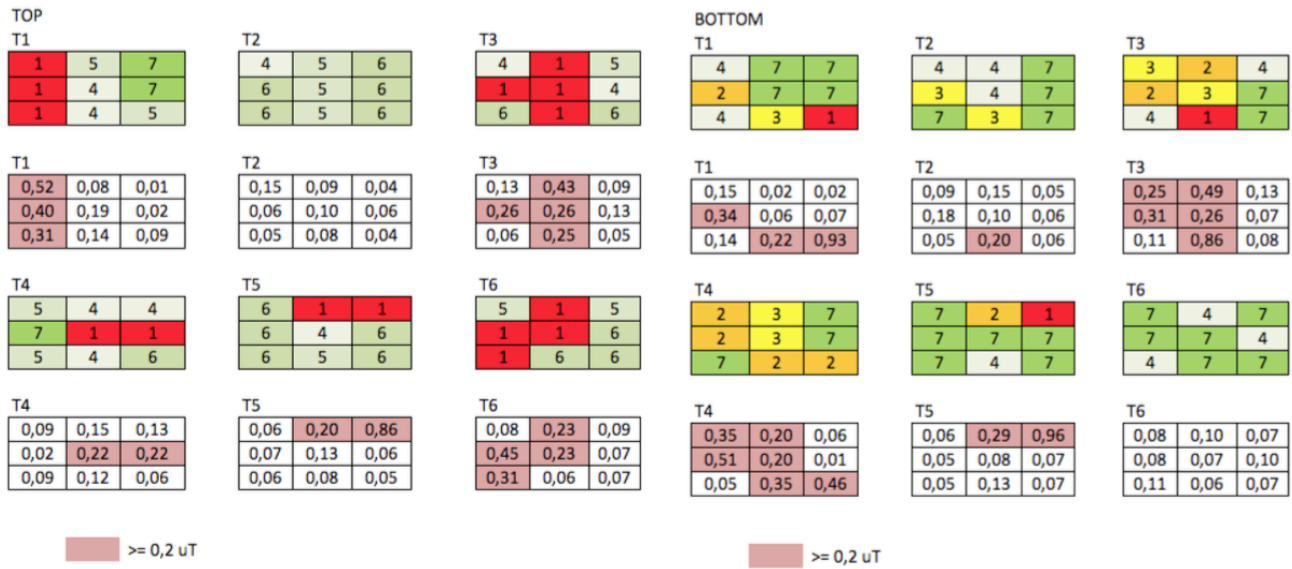

*Fig. 3. Dangerousness map on the 6 tablets*

## CONCLUSION

The paper considered the magnetic field radiation produced by tablets in the ELF range. The measurement process showed that tablet users can be exposed to dangerous levels of the magnetic field radiation. To classify these levels in meaningful ranges, the K-Medians clustering algorithm was used. The obtained levels were evaluated by taking into account the TCO proposed standard. At the end, the danger areas of the magnetic field radiation were established in order to warn the tablet users for possible risk assessments.

Future work will investigate the level of magnetic field radiation emitted from tablet when it operates under so-called stress condition, i.e. when tablet is intensively overloaded.